# The optical characteristics of the dust of sungrazing comet C/2012 S1 (ISON) observed at large heliocentric distances

Oleksandra Ivanova[a,b*], Volodymyr Reshetnyk[b,d], Yury Skorov[c], Jürgen Blum[c], Zuzana Seman Krišandová[a], Jan Svoreň[a], Pavlo Korsun[b], Viktor Afanasiev[e], Igor Luk'yanyk[f], Maxim Andreev[b,g,h]

[a] The Astronomical Institute of the Slovak Academy of Sciences
[*] Corresponding Author. E-mail address: oivanova@ta3.sk
[b-] Main Astronomical Observatory of NAS of Ukraine
[c-] University of Braunschweig, Institute for Geophysics and Extraterrestrial Physics
[d-] Astronomy and Space Physics Department, Faculty of Physics, Taras Shevchenko National University of Kyiv
[e-] Special Astrophysical Observatory of the Russian Academy of Science
[f-] Astronomical Observatory, Taras Shevchenko National University of Kyiv
[g-] Terskol Branch of the Institute of Astronomy of RAS
[h-] International Center for Astronomical, Medical and Ecological Research NAS of Ukraine

**Corresponding author:**

Dr. Oleksandra Ivanova
Astronomical Institute
of the Slovak Academy of Sciences,
Tatranska Lomnica
p.o. box 18, 05960
Slovak Republic

Phone: +421527879126
E-Mail address: oivanova@ta3.sk

**Highlights**

- We present the results of photometric observations of comet C/2012 S1 (ISON) performed from September 29 to December 16, 2012.
- We present the results of spectroscopic observations of comet C/2012 S1 (ISON) performed on February 9, 2013.
- Photometric colours of the comet are redder than solar colours.
- Analysis of spectral data shows the growth of the dust grain reflectivity with increasing wavelength.
- Molecular emissions were not detected in the spectra of this comet.
- The optical characteristics of the dust significantly depend on the chemical composition and structure. Effective medium approximations for light scattering by porous aggregates should be used for their evaluations.

**ABSTRACT**

We present an analysis of the photometric and spectroscopic data of the comet C/2012 S1 (ISON) observed at the heliocentric distances of 6.21 – 4.81 AU. The photometric observations were made with the 60-cm Zeiss-600 telescope (ICAMER, peak Terskol, Russia) and the spectroscopic observations were performed using the SCORPIO-2 focal reducer mounted in the prime focus of the 6-m BTA telescope (SAO RAS, Russia). We analyse the B, V and R-band images to describe the dusty cometary coma and to investigate its brightness, colours and dust production rate. The spectra cover the wavelength range of 3600–7070 Å. No emissions which are expected in this wavelength region were detected above the 3$\sigma$ level. The continuum shows a reddening effect with the normalized gradient of reflectivity along dispersion of 9.3 ± 1.1% per 1000 Å. A dust-loss rate was derived using the obtained values and under the different model assumptions. Our simulations clearly indicate that to retrieve dust production from the observational $Af\rho$ parameter is an ambiguous task. The result of such a procedure is strongly dependent on dynamical (e.g. effective density and cross-section) as well as optical (e.g. scattering coefficient and phase function) characteristics of dust grains. A variation of the mentioned parameters can lead to dramatic changes in the evaluation of mass production. We demonstrate that the dynamic and optical properties are interconnected via the microscopic properties of dust grains (effective size and porosity).

**Key words:** Comet C/2012 S1 (ISON); spectroscopy; photometry; colour; dust mass-loss rate;

## 1. Introduction

Comets were formed far away from the Sun and conserve the most primitive matter that exists up to date. Dynamically new comets are especially intriguing, as they enter the inner part of the Solar System for the first time, and their composition remains almost unchanged by solar radiation. Observations of new comets at different heliocentric distances, especially the comets that pass extremely close to the Sun at the perihelion (sungrazing comets), are very important for the verification of theoretical hypotheses about the structure of cometary nuclei and the physical processes that cause cometary activity at different distances from the Sun. One of such objects is a comet C/2012 S1 (ISON) (hereinafter referred to as C/2012 S1). This comet was discovered on September 21, 2012 with the 0.4-meter International Scientific Optical Network telescope (Green, 2012). This comet is a sungrazing comet with inclination $i$=62.4° and $e$=0.9999947 (Williams, 2013). Estimations of the effective nucleus radius of the comet were presented in various papers: Li et al. (2013) derived the 2-km nucleus size with the albedo $p$ = 0.04, Moreno et al. (2014) proposed 0.5 km with $\rho$=1000 kg·m$^{-3}$ and Scarmato (2014) gives approximately 0.8 km. Sekanina and Kracht (2014) argued a smaller size of the nucleus (it should not be more than 1 km across). From the Hubble Space Telescope observations of the comet, Lamy et al. (2014) obtained the upper limit of the cometary nucleus radius of about 0.68±0.02 km. Weaver et al. (2014) estimated the nucleus rotation period of about 10.4 hr from the cometary light curve.

The comet was discovered far from the Sun and it allowed one to monitor the comet's activity until it reached perihelion. The pre discovery point was at 9.4 AU (Nevski and Novichonok, 2012), and the comet was active. The comet showed outburst activity during the whole observation period at the large heliocentric distance and as well close to the Sun (Meech et al., 2013; Trigo-Rodríguez et al., 2013; Sekanina and Kracht, 2014). Brightness behaviour of the comet obtained from SOHO/STEREO is typical for small Kreutz comets (Knight and Battams, 2014; Jones et al., 2018). Usually, small Kreutz comets have tens of meters in radius or even a smaller size, and they are usually destroyed prior to reaching their perihelion (Iseli et al., 2002; Sekanina, 2003; Knight et al., 2010; Jones et al., 2018).

Comet C/2012 S1 passed perihelion on November 28, 2013 at a distance of 0.012567 AU. The comet did not survive while passing its perihelion, and a disintegration has likely begun prior to that point (Sekanina and Kracht, 2014; Curdt et al., 2014). In contrast to comet C/2012 S1, Kreutz Comet C/2011 W3 (Lovejoy) still survived for 1.6-days after its perihelion passage, despite the considerably smaller perihelion distance (0.0056 AU).

The comet C/2012 S1 was a “typical” comet to some extent. Many species were detected in the spectral data set of the comet including CN, $C_2$, $C_3$, CH, $NH_2$, [OI], NH, and OH (McKay et al., 2014). The comet was depleted in $C_2H_6$, $CH_3OH$, $CH_4$ molecules, typical in $C_2H_2$, HCN, and enhanced in $NH_2$ (DiSanti et al., 2014; Russo et al., 2016). DiSanti et al., (2014) observed the comet from 1.2 to 0.3 AU, and deferred that the CO abundance was pretty constant with the heliocentric distance. The abundances of the $H_2CO$ molecule increased with the decrease of heliocentric distance from depleted to enriched.

The photometric observations of the comet at the heliocentric distances from 9.3 AU to 0.44 AU were made at the optical, submillimeter and far-ultraviolet wavelengths (Meech et al., 2013, Trigo-Rodríguez et al., 2013; Weaver et al., 2014; Feldman et al., 2014; Lamy et al., 2014). The overall light curve of a comet C/2012 S1 based on about 300 selected ground-based observations obtained from 9.4 AU to 0.74 AU is presented in the manuscript of Sekanina and Kracht (2014). Detailed photometry was conducted for the comet in the period from October 2012 to November 2013 (Scarmato, 2014). The nucleus size, dust productivity, and rotation of the comet were estimated. The photometric observation of the comet with HTS broadband filters (Li et al., 2013) was obtained for a heliocentric distance of 4.15 AU. The parameter $Af\rho$ equal to 1340 and 1240 cm (with an aperture radius of 1.6") was estimated for the F606W and F438W filters respectively. Knight et al. (2015) observed the comet at heliocentric distances from 5.14 to 0.69 AU. Observations of the comet in the V band from the Swift showed the production rate *Afρ* from 750 to 861 cm at the heliocentric distance from 1.52 to 0.83, respectively (Bodewits et al., 2013).

The dust colours were slightly redder than those of the Sun with a slope near 6.0±0.2% per 1000 Å in the area around the nucleus (5 000 km) and >10% per 1000 Å in the area greater than 10,000 km down the tail (Li et al., 2013; Zubko et al., 2015).

Observations of Comet C/2012 S1 were conducted in the mid-infrared wavelength region (Ootsubo et al., 2014) when the comet was at the heliocentric distance $r$=1.28 AU. The spectral analysis showed the presence of small amorphous olivine grains and did not reveal any clear crystalline silicate feature presence. Wooden et al. (2014) presented thermal models for Comet C/2012 S1, when the comet was at the heliocentric distance $r$= 1.15 AU, and found that micron-sized grains with fractal dimension of $D$=2.7 dominated in the cometary coma. The authors mentioned that the grain size of the comet is more similar to the Oort cloud Comets C/2007 N4 (Lulin) (Woodward et al., 2011) and C/2006 P1 (McNaught) (Kelley et al., 2007).

Keane et al. (2014) presented results of submillimeter dust-continuum observations of the comet obtained before perihelion at the heliocentric distance $r$ = 0.0125 AU. Analysis of the observations allowed one to detect the 1-mm-sized dust particles in the cometary coma and, subsequently, their large-scale fragmentation.

Monte Carlo modelling of the cometary dust tail in the post-perihelion (Moreno et al., 2014) showed that dust was very small in size, ranging from 0.1 to 50 μm in radius, distributed according to a power law with an index of –3.5.

In this paper we present the analysis of optical observations of Comet C/2012 S1 at a pre-perihelion distance from 6.22 AU to 4.81 AU, respectively. The paper is organized as follows: observations and data reduction are presented in Section 2. Section 3 introduces description of the model parameters and the outcomes of image analysis, estimation of magnitude, colour, dust production based on the *Afρ* technique. The main results and their discussion are presented in Section 4.

## 2. Observations and data reduction

Comet C/2012 S1 (ISON) was observed with the 60-cm Zeiss 600 Telescope of the peak Terskol Observatory (IC AMER) during several sets from September 29 to December 16, 2012. We observed the comet when it was at the heliocentric distances from 6.22 AU to 4.81 AU, respectively.

The CCD PixelVision Vienna camera was used as a detector. The dimension of the image was 1024 x 1024 pixels. The full field of view of the CCD was 10.7 x 10.7 arc minutes, and the image scale was 0.63 arc seconds per pixel. Photometric images of the comet were obtained in the Johnson B, V, and R broadband filters centred at 4330 Å, 5450 Å, and 6460 Å respectively.

The 2×2 binning was applied during the observations. Air masses varied from 1.44 to 1.63. The images of SA52-193 (Landolt, 1992) were obtained for the absolute calibration of the comet images. All the images were bias-subtracted and flat-field corrected. We used the programs of the IDL library (Goddard Space Flight Center) to calculate the sky background (Landsman, 1993). All the images were shifted to the same centre of the comet (for each filter in the observation set), and stacked together to increase the S/N ratio.

All of the nights were photometric**,** the seeing value measured as the average FWHM of several sample stars was about 1.8 arcsec during our observations. The residual sky background was estimated with the use of an annular aperture. The brightness variability of the comet and standard stars in the R filter are shown in Figure 1.

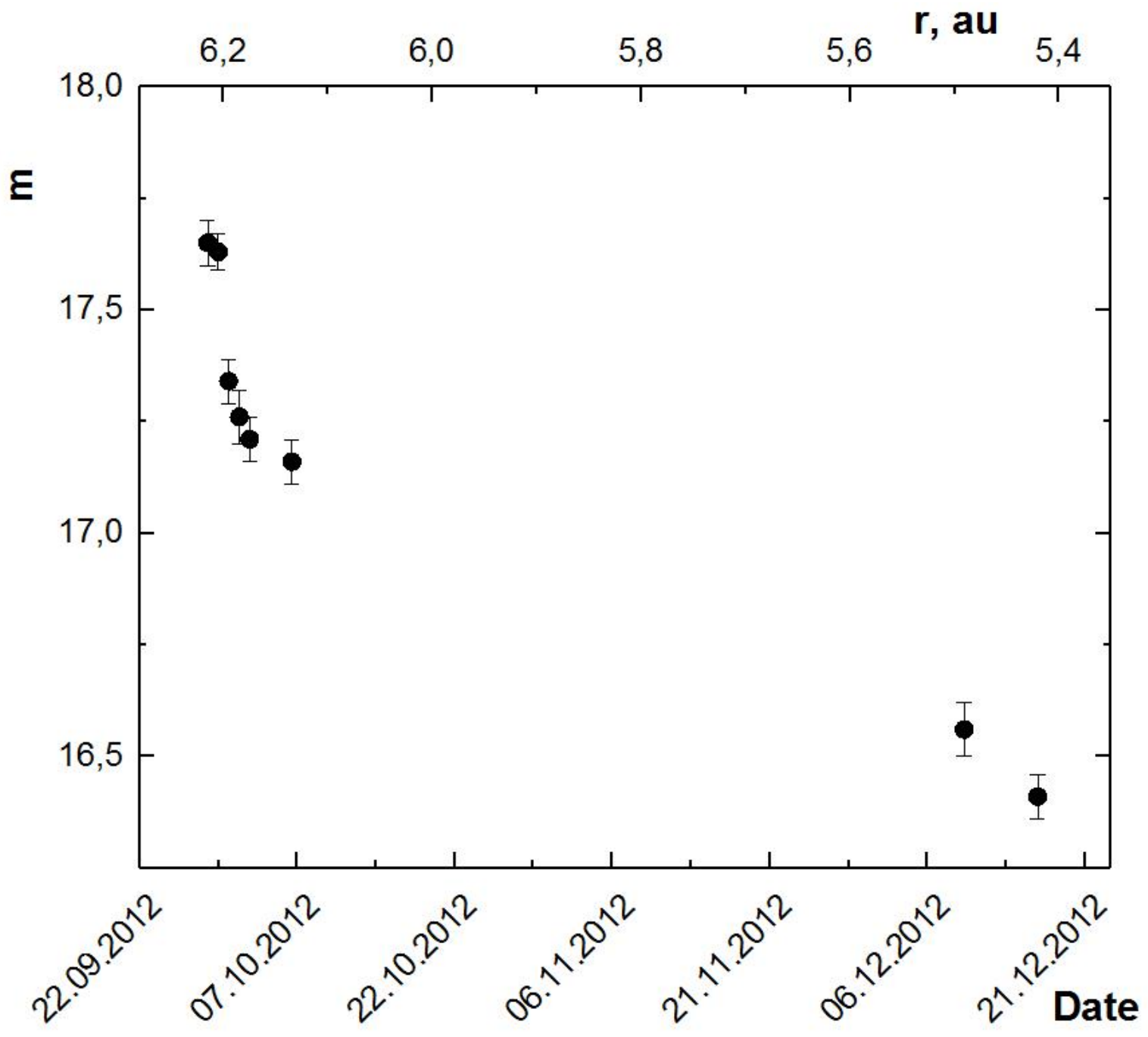

**Fig. 1.** Brightness of the comet in R broadband filter as a function of the heliocentric distance *r* and observation time *t*.

The spectroscopic observations of C/2012 S1 were carried out at the SAO RAS (Special Astrophysical Observatory of Russian Academy of Sciences, Russia) 6-m BTA telescope on February 09, 2013, when the heliocentric and geocentric distances of the comet were 4.81 and 3.98 AU, respectively. The SCORPIO-2 focal reducer mounted at the prime focus of the telescope was operated in long-slit spectroscopic modes (Afanasiev and Moiseev, 2011). An E2V-42-90 CCD sized 2K × 4K was used as a detector. The size of one pixel is 16 ×16 μm which corresponds to 0.18″× 0.18″ on the sky plane.

The spectroscopic observations were conducted with a long-slit mask. The height and the width of the slit were 6.1″ and 1″ respectively. The VPHG1200@540 transparent grizm was used as a disperser in the spectroscopic mode. The spectra cover the wavelength range of 3600–7070 Å and the spectral resolution, which was defined by the width of the slit, was about 5 Å. The spectroscopic images were binned along the spatial direction as 2×4. A lamp with a smoothly varying energy distribution was used to compensate for the non-uniform sensitivity of the CCD pixels in the spectroscopic mode. Wavelength calibrations were made exposing a He–Ne–Ar-filled lamp. The spectrum of the morning sky was exposed to estimate variations of the background night sky spectrum along the spatial direction. For the spectral calibration of the spectra, the spectrophotometric standard BD+75d325 was observed (Oke, 1990). The spectral behaviour of the atmospheric extinction was also taken from Kartasheva and Chunakova (1978). Standard reduction procedures for the obtained spectroscopic data were performed. We used the Scorpio_2x4K.lib package operating under the IDL to remove biases from the observed frames, to clean the frames from cosmic events, and to correct their geometry. The background night sky spectrum was removed using the expositions of the morning sky spectrum. The latter was weighted in order to fit the observed level of the night sky spectrum in each column along the slit. Detailed information about the observations is presented in Table 1 and Fig. 2.

**Table.1.** Observation log of comet C/2012 S1 (ISON)

| Data, UT | Exp.[a] | $T^b$,s | r, AU | Δ, AU | $\alpha^c$ | $PA_{sun}^d$ | $PA_v^e$ | $m^f$ | Mode |
|---|---|---|---|---|---|---|---|---|---|
| **Sept. 29.0130, 2012** | 15 | 1350 | 6.225 | 6.573 | 8.4 | 286.5 | 91.4 | 17.65±0.05 | R |
| **Sept. 29.0408, 2012** | 15 | 1350 | | | | | | 18.87±0.09 | B |
| **Sept. 29.0526, 2012** | 15 | 1350 | | | | | | 18.20±0.09 | V |
| | | | | | | | | | |
| **Sept. 29.9863, 2012** | 15 | 1350 | 6.216 | 6.548 | 8.5 | 286.4 | 90.9 | 17.63±0.05 | R |
| **Sept. 30.0022, 2012** | 21 | 1890 | | | | | | 18.85±0.10 | B |
| **Sept. 30.0245, 2012** | 21 | 1890 | | | | | | 18.17±0.09 | V |
| | | | | | | | | | |
| **Oct. 1.0249, 2012** | 14 | 1170 | 6.206 | 6.522 | 8.6 | 286.3 | 90.5 | 17.34±0.05 | R |
| **Oct. 1.0397, 2012** | 17 | 1530 | | | | | | 18.59±0.08 | B |
| **Oct. 1.0601, 2012** | 13 | 1260 | | | | | | 17.91±0.08 | V |
| | | | | | | | | | |
| **Oct. 2.01547, 2012** | 15 | 1350 | 6.196 | 6.497 | 8.6 | 286.1 | 90.0 | 17.26±0.05 | R |
| **Oct. 2.03135, 2012** | 19 | 1710 | | | | | | 18.51±0.09 | B |
| **Oct. 2.05156, 2012** | 17 | 1530 | | | | | | 17.83±0.08 | V |
| | | | | | | | | | |
| **Oct. 3.02524, 2012** | 13 | 1170 | 6.186 | 6.471 | 8.7 | 286.0 | 89.5 | 17.21±0.05 | R |
| **Oct. 3.03966, 2012** | 15 | 1350 | | | | | | 18.45±0.09 | B |
| **Oct. 3.05552, 2012** | 13 | 1170 | | | | | | 17.78±0.08 | V |
| | | | | | | | | | |
| **Oct. 7.07503, 2012** | 10 | 900 | 6.142 | 6.368 | 8.9 | 285.5 | 87.0 | 17.16±0.05 | R |
| **Oct. 7.05059, 2012** | 13 | 1170 | | | | | | 18.40±0.08 | B |
| **Oct. 7.06443, 2012** | 10 | 900 | | | | | | 17.73±0.08 | V |
| | | | | | | | | | |
| **Dec. 10.9495, 2012** | 22 | 1980 | 5.492 | 4.709 | 6.7 | 271.5 | 289.7 | 16.56±0.05 | R |
| **Dec. 10.9612, 2012** | 22 | 1980 | | | | | | 17.82±0.08 | B |
| **Dec. 10.9729, 2012** | 22 | 1980 | | | | | | 17.14±0.08 | V |
| | | | | | | | | | |
| **Dec. 17.0833, 2012** | 22 | 990 | 5.430 | 4.586 | 5.8 | 271.0 | 287.5 | 16.41±0.05 | R |
| **Dec. 17.1004, 2012** | 22 | 990 | | | | | | 17.65±0.08 | B |
| **Dec. 17.1162, 2012** | 22 | 990 | | | | | | 16.98±0.08 | V |
| | | | | | | | | | |
| **Feb. 9.6869, 2013** | 5 | 180 | 4.805 | 3.977 | 7.1 | 109.0 | 269.7 | - | spectra VPHG 1200@540 |

a- Number of exposures
b- Total integration time
c- Phase angle, degree
d- Position angle of the extended Sun - target radius vector, degree
e- Position angle of the velocity vector, degree
f- Obtained with an aperture radius of 5 arcsec

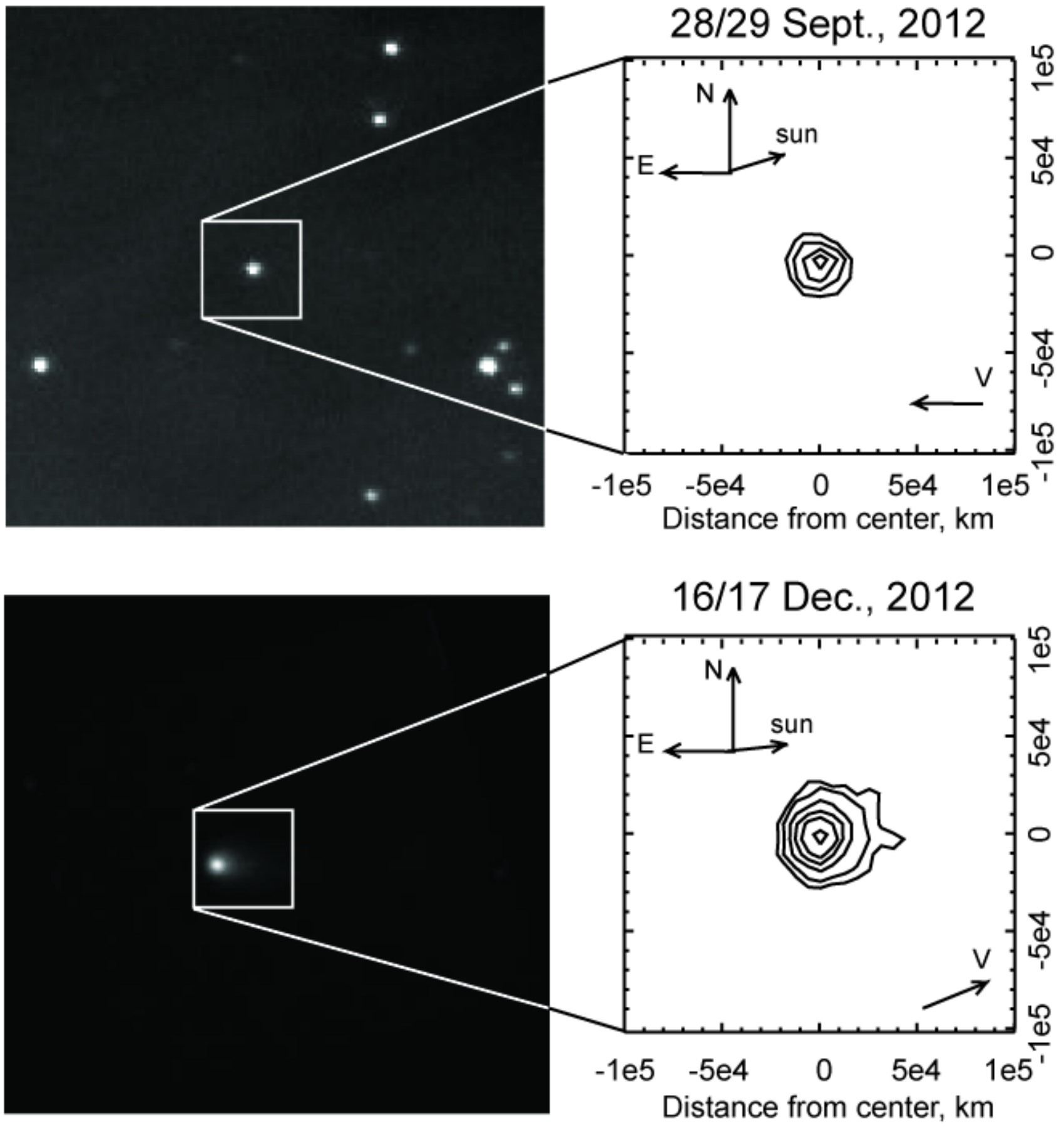

**Fig. 2.** Co-added images of Comet C/2012 S1 (ISON) observed through the broad band R filter together with relative isophotes differing by a factor √2 for September 29, 2012 (top) and December 16, 2012 (bottom) respectively. Celestial north, east, the motion V, and sunward directions are denoted.

## 3. Analysis of the spectra

We recorded five two-dimensional spectra of the comet when the comet was closer to the Sun (see Table 1). To increase the S/N ratio of the data, we co-added the available spectra. The derived behaviour of the energy in the spectrum of C/2012 S1 along dispersion is displayed in Fig. 3(a).

A scaled solar spectrum (Neckel and Labs, 1984) convoluted to the cometary spectrum resolution and corrected for the reddening effect is also superimposed. To check the spectrum on the presence of possible molecular emissions, we subtracted the fitted continuum from the observed spectrum. The result is depicted in Fig. 3(b). The S/N ratio of the residuals is a strong function of wavelength. Therefore, $\pm 3\sigma$ levels have been calculated for the spectral windows, where the strongest emissions of the $C_2$ and $CO^+$ molecules are expected, and are shown with dotted lines in the figure. There are no signals above these thresholds in the above-mentioned regions. We hence state that at least the emissions of $C_2$ (the head of the strongest band (0-0) is located at around 5165 Å) and $CO^+$ (the strongest band (2,0) is located within the 4250–4280 Å wavelength region) above the $3\sigma$ level are not detected in the spectrum.

We calculated upper limits of the emission fluxes and upper limits of the production rates of the neutrals, which could be detected at large heliocentric distances. In order to calculate the relative fluxes we used an approach similar to that presented by Rousselot et al. (2014). The upper limits of the production rates of the neutrals were derived using the Haser model (Haser, 1957). The model parameters for the neutrals were taken from Korsun et al. (2016). G-factor for CN, which depends on the heliocentric velocity of the comet, was taken from Schleicher (2010). The resulting upper limits are presented in Table 2.

**Table 2. Upper limits for the main cometary species not detected in the spectra.**

| Molecule | Central wavelength, Å/Δλ | Amplitude × $10^{-16}$ erg $s^{-1}$ $cm^{-2}$ $Å^{-1}$ | Flux × $10^{-16}$, erg $s^{-1}$ $cm^{-2}$ | Gas production rate, $10^{25}$ mol $s^{-1}$ |
|---|---|---|---|---|
| CN | 3870/62 | <7.24 | <42.83 | <1.4 |
| $C_3$ | 4062/62 | <4.00 | <23.66 | <0.1 |
| $CO^+$ | 4266/64 | <3.53 | <20.88 | |
| $C_2$ | 5141/118 | <1.76 | <10.41 | <0.8 |
| $NH_2$ | 5715/90 | <1.43 | <8.46 | <3.5 |

We have also examined variations of the reflectivity *S(λ)* along dispersion, which is expressed through the division of the cometary spectrum by the scaled solar spectrum. We consider that the first degree of the polynomial fitting, which we used to fit the cometary continuum is adequate, as it is shown in Fig. 3(c). The result demonstrates the growth of dust grain reflectivity with increasing wavelength and can be fixed quantitatively, as the normalized gradient of reflectivity $S'(\lambda_1, \lambda_2)$ is expressed in percent per 1000 Å. The measurements were obtained at the wavelengths $\lambda_1$ and $\lambda_2$ (in Å), where $\lambda_1 > \lambda_2$. The derived reddening is equal to 9.3%± 1.1% per 1000 Å in the range of 3800-5400 Å and 2.5% ± 1.2% for 5400—6800 Å, respectively.

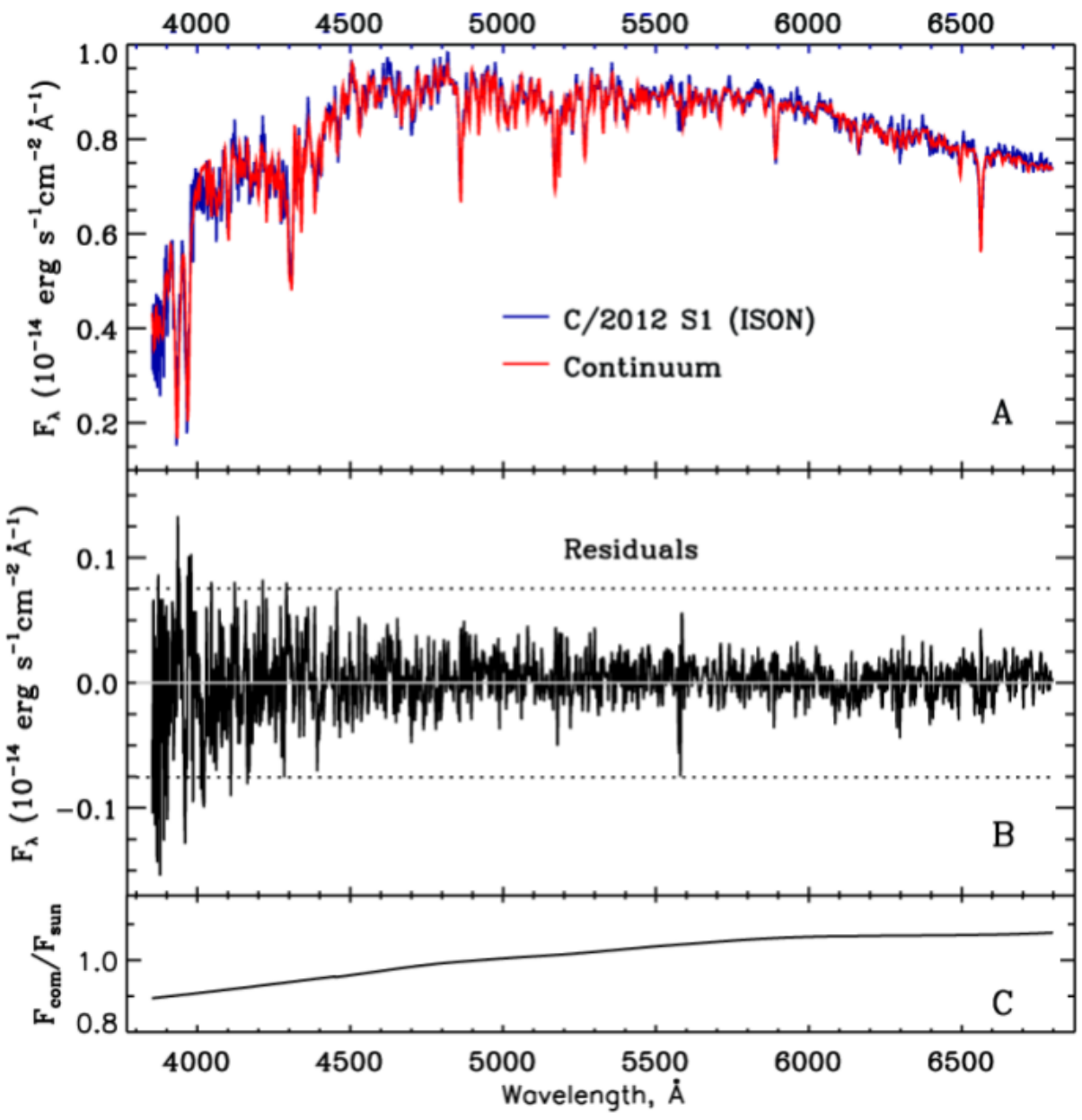

**Fig. 3.** The spectrum of Comet C/2012 s1 (ISON) with the scaled solar spectrum (a), the observed spectrum fitted to continuum residuals, dotted lines indicate ±3σ levels, where emissions of $C_2$ and $CO^+$ are expected (b), and the polynomial fitting of the ratio of cometary spectrum to solar spectrum (c).

## 4. Image analysis

### 4.1 Morphology of the cometary dust coma

Our observations of the object in September and October 2012 showed low cometary activity. Comet C/2012 S1 (ISON) in our images, obtained with the 60-cm telescope before December 2012, has a star–like shape, no extended coma and tail. The diffraction-limited resolution of our telescope is about 0.2 arcsec (in the R filter), spatial resolution is 0.6 arcsec pixel$^{-1}$. The seeing was 1.4-1.8 arcsec. The isophote contours of the image of Comet C/2012 S1 (ISON) taken on September 29 and December 16, 2012 with the R filter are shown in Fig. 2. One can see that the comet has a small coma or tail only in December. Although Meech et al. (2013) state that from September 2012 and on the comet possessed a visible coma.

The mean profile was built as a result of averaging of all the obtained azimuth profiles (Fig. 4). The profiles of the star and comet, obtained on September 29, 2012 are very similar. The reference star FWHM is about 1.8 arcsec and the comet image FWHM is 2 arcsec.

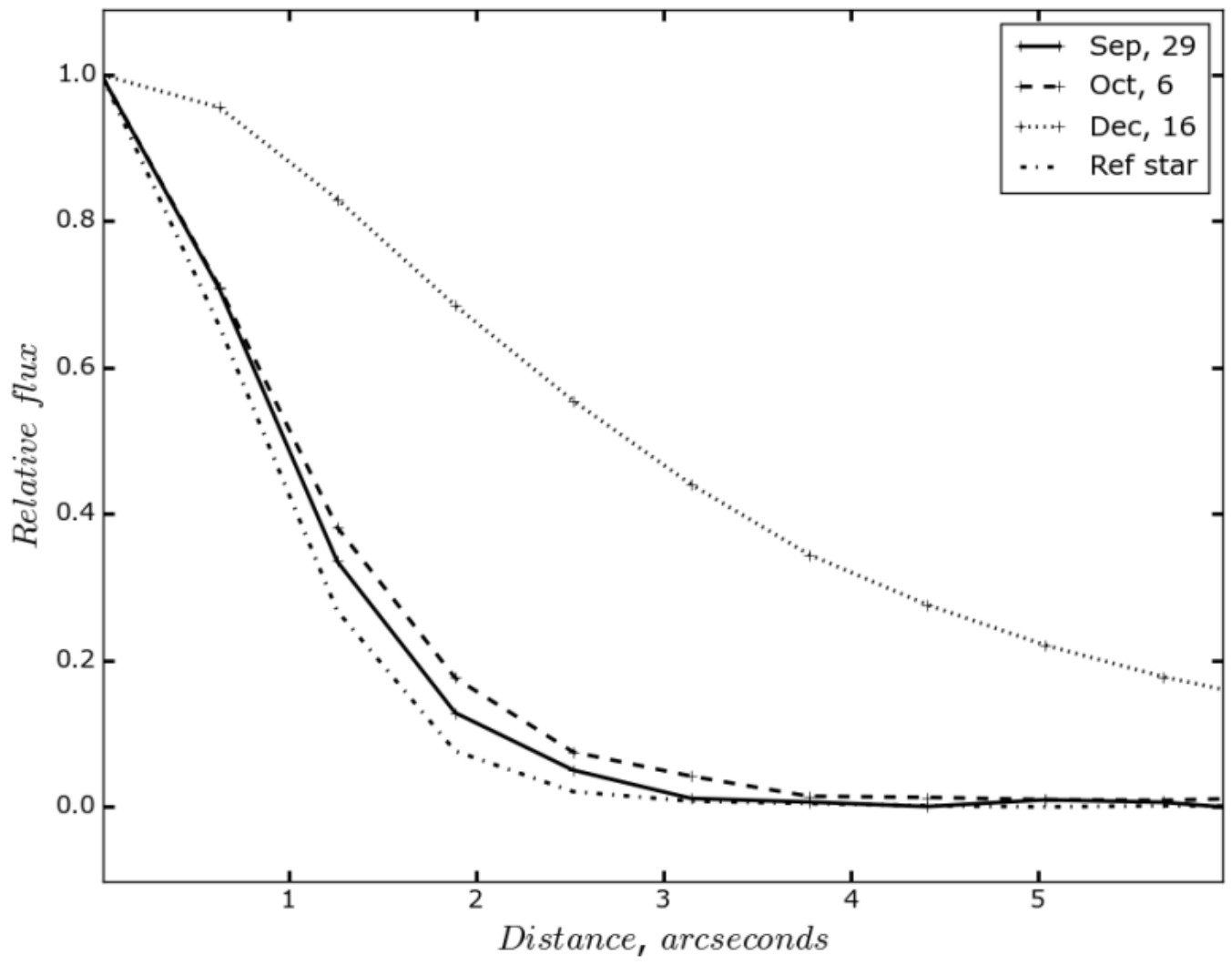

**Fig. 4.** Azimuthally averaged radial profiles of the image of Comet C/2012 S1 and reference star (the R filter).

From Fig. 4, one can see that the brightness of C/2012 S1 was slowly increasing during September – October 2012. In December 2012, the brightness of the comet was a few times greater than in the previous months. During the observations from September to October, the comet practically did not change its own shape and stayed star-like or showed a small tail in December only.

We can conclude that during the autumn 2012, Comet C/2012 S1 had a star-like image in small-telescope observations. However, the detailed analysis with radial profile distribution shows that in September 2012 the comet had a weak coma and possibly a small tail which grew during two months. The comet tail observations confirm the results of the other authors (Meech et al., 2013). Based on our images of the comet we can only study the overall shape, of the coma without detailed analysing, how was maked in Samarasinha et al., (2015). Authors (Samarasinha et al., 2015) presented the results of a global coma morphology campaign for comet C/2012 S1 (ISON), which received images at heliocentric distance from 5.1 AU to 0.35 AU, when the comet was closer to the Sun.

## 4.2 Colour excess and reflectivity gradient of the cometary dust

We used the images obtained with the broadband B, V, and R filters to measure the magnitude of the nucleus of Comet C/2011 S1 (ISON). We also observed the standard stars field (SA52-193) at different airmasses to provide information for the photometric reduction (Landolt, 1992). The cometary magnitude is given by

$$m_c = -2.5log\left[\frac{I_c(\lambda)}{I_s(\lambda)}\right] + m_{st} - 2.5logP(\lambda)\Delta M \qquad (1)$$

where $I_s$ and $I_c$ are the measured fluxes of the star and the comet in counts, respectively, $m_{st}$ is the magnitude of the standard star, $P$ is the sky transparency that depends on the wavelength, $\Delta M$ is the difference between the air masses of the comet and star. The difference between the comet and standard field airmasses was considered to be small and was compensated for with the mean monochromatic extinction coefficients measured for Peak Terskol Observatory, which is located in the same mountain region (Kulyk et al. 2004).

Using equation (1), the magnitude of the comet was calculated for an aperture radius of 5 arcsec (from 23837 km to 16630 km). All corrections for compensation of time-variable or heliocentric distance effects were made. The reduced magnitude can be computed as $m_c - 2.5k\log(\Delta) - 2.5n log(r)$ where the parameters $n$ and $k$ taken from http://www.isoncampaign.org/Present-archive-nov11. We obtained the comet absolute magnitude for all sets of our observations. The results are presented in Table 2.

We calculated the colours of the cometary dust and the colour excess of the dust - $\Delta m = \Delta m_c - \Delta m_s$, where $\Delta m_c$ is the comet's colour and $\Delta m_s$ is the Sun's colour. The solar colour indices $B-V$ =+0.65 and $V-R$=+0.52 (Allen, 1976, Lamy et al., 1988) are in the Johnson filter system. The colours of the comet show that the scattered light is redder than that of the Sun. The results are given in Table 3.

Also, we calculate the reflectivity gradient $S^{\prime}$ (it shows the wavelength dependence of the light scattered by the dust), which is the percentage change in the strength of the continuum per 1000 Å (Jockers, 1997; Meech, et al., 2009):

$$S^{\prime} = \frac{2}{\Delta\lambda}\frac{(10^{0.4\cdot\Delta m}-1)}{(10^{0.4\cdot\Delta m}+1)} \times 10^5, \qquad (2)$$

where $\Delta m$ is the comet colour minus the Sun colour, $\Delta\lambda = \lambda_1 - \lambda_2$ is the difference in effective wavelengths. Calculations of variations of the dust colour and reflectivity gradient are important for diagnosing the properties of dust of the comets. The values $S^{\prime}$ in different filters are given in Table 3.

**Table.3**. Dust colour and reflectivity gradient of Comet C/2012 S1 (ISON)

| **Data** | **r, AU** | **Δ, AU** | **Phase angle, degree** | **Filter/ range, Å** | **$\Delta m_c$** | $\Delta m_c - \Delta m_s$ | **$S^{\prime}$, % per 1000Å** | **Mode** |
|---|---|---|---|---|---|---|---|---|
| 28.09.2012 | 6.225 | 6.573 | 8.4 | B-V | 0.67±0.13 | 0.02 | 1.72±0.7 | Image |
| | | | | V-R | 0.55±0.10 | 0.03 | 2.74±1.1 | Image |
| 29.09.2012 | 6.216 | 6.548 | 8.5 | B-V | 0.68±0.13 | 0.03 | 2.58±0.9 | Image |
| | | | | V-R | 0.54±0.10 | 0.02 | 1.82±0.8 | Image |
| 30.09.2012 | 6.206 | 6.522 | 8.6 | B-V | 0.68±0.11 | 0.03 | 2.58±0.9 | Image |
| | | | | V-R | 0.57±0.09 | 0.05 | 4.56±1.4 | Image |
| 1.10.2012 | 6.196 | 6.497 | 8.6 | B-V | 0.68±0.12 | 0.03 | 2.58±0.9 | Image |
| | | | | V-R | 0.57±0.09 | 0.05 | 4.56±1.4 | Image |
| 2.10.2012 | 6.186 | 6.471 | 8.7 | B-V | 0.67±0.12 | 0.02 | 1.72±0.7 | Image |
| | | | | V-R | 0.57±0.09 | 0.05 | 4.56±1.3 | Image |
| 6.10.2012 | 6.142 | 6.368 | 8.9 | B-V | 0.67±0.11 | 0.02 | 1.72±0.7 | Image |
| | | | | V-R | 0.57±0.09 | 0.05 | 4.56±1.4 | Image |
| 10.12.2012 | 5.492 | 4.709 | 6.7 | B-V | 0.68±0.11 | 0.03 | 2.58±0.9 | Image |
| | | | | V-R | 0.58±0.09 | 0.06 | 5.47±1.6 | Image |
| 16.12.2012 | 5.430 | 4.586 | 5.8 | B-V | 0.67±0.11 | 0.02 | 1.72±0.7 | Image |
| | | | | V-R | 0.57±0.09 | 0.05 | 4.56±1.3 | Image |
| 9.02.2013 | 4.805 | 3.977 | 7.1 | 3800-5400 | - | - | 9.3%± 1.1% | Spectra |
| 9.02.2013 | 4.805 | 3.977 | 7.1 | 5400-6800 | - | - | 2.5% ± 1.2% | Spectra |
| Average colour of active long-period comets at large heliocentric distances: B-V=0.78±0.02 V-R=0.47±0.02 (Jewitt, 2015) and pre-perihelion V=0.82±0.02 V-R=0.43±0.04 (Kulyk et al., 2018); | | | | | | | | |

Table 3 shows the variations of the colour index and reflectivity gradient of the comet from 1.72 to 5.47 % per 1000 Å. The obtained results are close to the results found for the near-nucleus area for this comet at a heliocentric distance of ~ 4 AU (Li et al., 2013; Zubko et al., 2015). The difference in the colour slope inferred from the spectrum and those results from our photometry can be caused by change in dust-particle population at different heliocentric distances. At the closer solar distance (4.8 AU) the outflow of gas can be somewhat stronger compared to the larger solar distance (6.2 AU). A weaker outflow of gas should expel in general smaller dust particles at r=6.2 AU as compared to r=4.8 AU. It is important to stress that small particles tend to have a bluer photometric colour that is qualitatively consistent with the observed difference in colour slope.

As one can see from Table 2 and Fig. 5 values of the colour and reflectivity gradient a little increase at the heliocentric distance from 6.21 to 6.18 AU that can be probably indicative of a slow increase in activity in this period, which described and analysed in Meech et al. (2013).

The character of this variation with a phase angle is close to the results presented for Comet C/1975 V1 (West) in the paper by Zubko et al., (2014). This variation is likely to be independent from the changes of dust properties in the period of observations and is related to the geometry of observations.

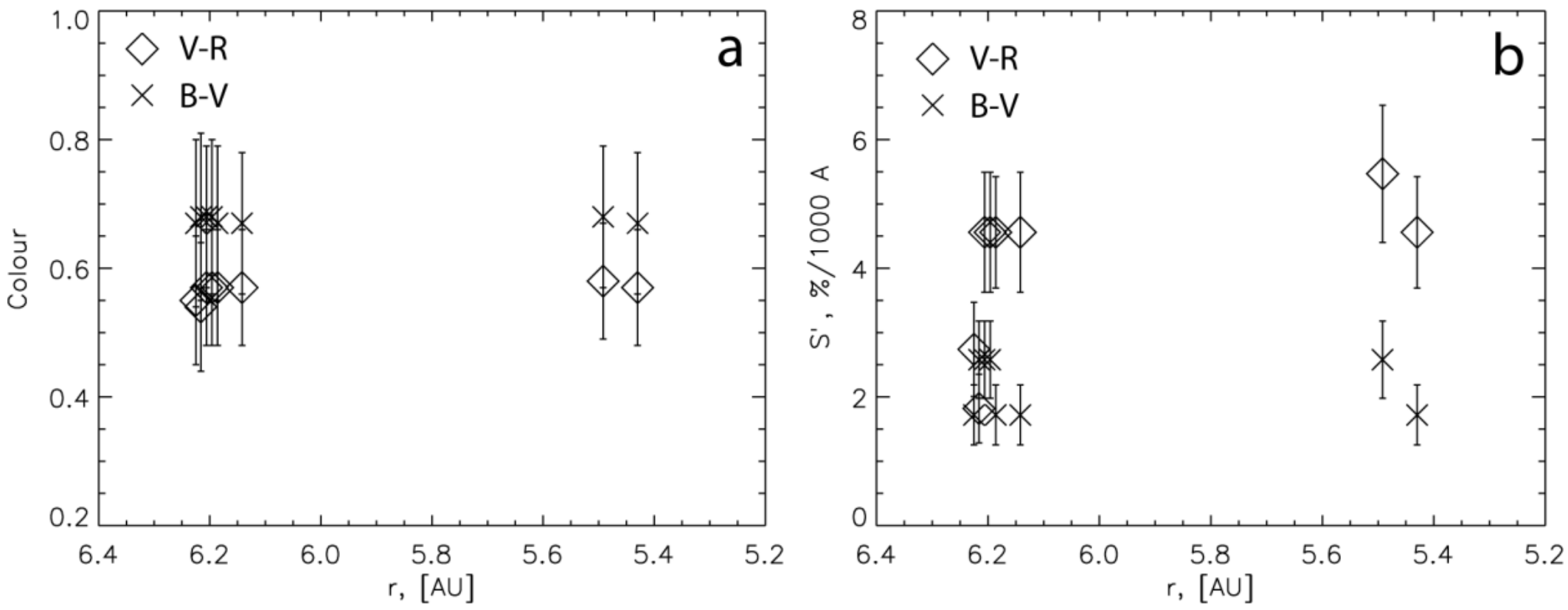

**Fig.5.** The colour (a) and colour slope (b) in the coma of C/2012 S1 (ISON). Average colour of active long-period comets at large heliocentric distances: B-V=0.64±0.02; V-R=0.35±0.01 (Jewitt, 2015)

### 4.3 Afρ parameter calculation

We observed the comet C/2012 S1 (ISON) beyond the Jupiter's orbit (Table 1). No appreciable gas emissions were detected in the visible spectra at these distances. This is not surprising since the spectra of new comets usually contain only a dust continuum spectrum without significant gas emissions at such large heliocentric distances. Only two comets (29P/Schawasmann-Wachmann 1 and C/2002 VQ94 (LINEAR)) displayed gas emissions CN, $CO^+$ and $N_2^+$ in the spectra at these heliocentric distances (Cochran et al., 1980; Korsun et al, 2006; 2008; Ivanova et al., 2016), but these emissions are outside the R broadband filter, which was used in our observations. Because the colour indexes of the comet are redder than the solar ones (Sec. 3.2), we assume that the $C_2$ emission does not influence the dust colour. Thus, under the above conditions, the data obtained with broadband R filters allow us to evaluate the total dust productivity of the comet.

We begin our study with an analysis of the parameter $Af\rho$, which is widely used to estimate the total dust production. Its popularity is owing to its independence of the observational equipment, the wavelength of images taken and the geometric arrangement of the observations. A'Hearn et al. (1984) provided the definition of the $Af\rho$ parameter as a product of the grain's albedo $A$ multiplied by the filling factor $f$ and the radius of the coma under investigation $\rho$. The filling factor is defined as a ratio of the particle total cross section to the projected field of view, and it is related to the total optical density of the coma. Thus, the $Af\rho$ can be interpreted as the amount of solar radiation reflected from the dust coma.

We calculate the $Af\rho$ parameter for comet C/2012 S1 for different distances (Table 4) according to the classical A'Hearn approach.

**Table.4.** $Af\rho$ **values (expressed in cm), computed from our photometric data.**

| Data | r, [AU] | Δ, [AU] | Phase angle, [degree] | Afρ*, [cm] |
|---|---|---|---|---|
| 28.09.2012 | 6.225 | 6.573 | 8.4 | 1137±148 |
| 29.09.2012 | 6.216 | 6.548 | 8.5 | 1151±150 |
| 30.09.2012 | 6.206 | 6.522 | 8.6 | 1492±192 |
| 1.10.2012 | 6.196 | 6.497 | 8.6 | 1595±199 |
| 2.10.2012 | 6.186 | 6.471 | 8.7 | 1658±207 |
| 6.10.2012 | 6.142 | 6.368 | 8.9 | 1688±211 |
| 10.12.2012 | 5.492 | 4.709 | 6.7 | 1731±171 |
| 16.12.2012 | 5.430 | 4.586 | 5.8 | 1892±183 |

*we used the cometary magnitude obtained in the R filter

### 4.4 Dust mass loss of the comet

If $Af\rho$ is known, one can estimate a total dust production from the comet. Recently an extended review and analysis of this procedure was presented by Fink and Rubin (2012). In this article the authors re-introduced the parameter $Af\rho$ using individual dust characteristics assuming that the grain's albedo $A$ is the scattering efficiency of the dust particle $Q_{sca}$ multiplied by the corresponding phase function $p$. Thus, they wrote:

$$Af\rho = Q_{sca} p f \rho \tag{3}$$

In turn the filling factor of dust grains can be written as:

$$f = \frac{N\sigma}{\pi\rho^2}, \tag{4}$$

where $N$ is the number of dust particles and $\sigma$ is the particle geometrical cross section. Obviously, the changes of the phase function $p$, the scattering efficiency $Q_{sca}$ and the grain size can dramatically change the estimation of dust number density $N$ in the cometary coma. Hereafter we develop this approach and analyse uncertainties arising when the dust production is derived from the observed $Af\rho$.

The dust production [particles per second] is determined by the equation

$$N_d = \frac{Af\rho}{\sigma Q_{sca} p} 2 v_d, \tag{5}$$

where $v_d$ is the terminal dust velocity, $Q_{sca}$ is the scattering efficiency, and $p$ is the phase function. A neat derivation of this equation is presented in (Fink and Rubin, 2012). The parameter $Af\rho$ is derived from our observations according to the A'Hearn's approach (see Table 4 above). For simplicity, we examine a monodisperse dust, i.e. an effective radius of the quasi-sperical dust grains $R_{eff}$ (see Appendix) is fixed for each model implementation. To estimate the dust terminal velocity, the equation of grain motion is solved

$$\ddot{\vec{r}} = G\frac{M}{r^3}\vec{r} + \frac{1}{2} C_D \frac{\sigma}{m_d} n_{mol} m_{mol} |\vec{u} - \vec{v_d}| (\vec{u} - \vec{v_d}) \tag{6}$$

where the first term corresponds to the cometary gravity force (the nucleus mass $M$ is equal to $6.5 \cdot 10^{11}$ kg, estimated from radius of the nucleus 0.68 ± 0.02 km (Lamy et al., 2014) and average density **0.5 g/cm$^3$ (Kinght and Walsh, 2013**), and the second one is the gas drag force, which can be expressed as a product of the drag coefficient $C_D$ evaluated for a spherical particle, the ratio of $\sigma$ to the mass of the dust particle $m_d$, the mass $m_{mol}$, and the number density $n_{mol}$ of CO gas molecules, and the relative squared velocity (for details see, e.g., Skorov et al. 2016). The gravity of the cometary nucleus reduces the dust acceleration by expanding gas. At a distance of several dozen kilometres the gas and dust are decoupled, and the speed of dust grains reaches the terminal value. The terminal velocity is higher for the small and/or more porous particles, because for the considered model porous aggregates the larger the cross-section-to-mass ratio $\frac{\sigma}{m_d}$, the higher the terminal velocity $v_d$ .

Based on the calculated $v_d$ and optical properties of the carbon and silicate dust aggregates, dust production can be straightforwardly calculated. Two types of porous aggregates (Ballistic Agglomeration - BA and Ballistic Agglomeration with 2 Migrations - BAM2) are examined based on the combination of the Mie model and the effective medium theory (EMT) with the Maxwell-Garnett mixing rule. The model of the optical properties of these porous grains is presented in Appendix.

Hereafter we use only one value $Af\rho$ corresponding to the observations on 29.09.2012. The corresponding scattering phase angle is about 8.5°. We do this in order to clearly show how the retrieved dust production depends on both dynamical (e.g. effective density and cross-section) and optical (e.g. scattering coefficient and phase function) characteristics of dust aggregate. The results of such simulations for the carbon and silicate aggregates are shown in Fig. 6 in the left and right columns, respectively. The modelling data obtained for different monomer sizes are plotted in the different rows.

As expected the terminal velocity of porous aggregates is higher than that of a sphere of the same mass (by about 50%) and decreases with mass (by about 50—60%) due to the decreasing cross-section-to-mass ratio. Which leads to an increase in the dust production $M_d$ calculated by formula (5). However, the change due to this cause (i.e. the change in grain structure) is only a small part of the total change due to variations in the optical characteristics. Thus, for carbon particles $M_d$ may vary about ten times in the considered mass interval. For these grains having a noticeable imaginary part of the refractive index, the change of dust production on the mass of the aggregate is determined mainly by the phase function variations. Indeed, the scattering efficiency $Q_{sca}$ varies insignificantly with a change in the particle mass, and weakly depends on the type of aggregate (see the left column of Fig. A1). We note that the dust production is sensitive to the aggregate porosity: the curves obtained for the BA and BAM1 differ by 4—5 times and this difference is close to the constant in the examined mass interval. The influence of the monomer size is not so strong. Dust production evaluated for the particles of the same mass (~$2\cdot10^{-13}$ kg) constructed from the monomers whose sizes are different by an order of magnitude (top and bottom rows) differs only twice. Thus, systematic simulations show that for the opaque aggregates (carbon particles in our work) 1) the variations of the size (or mass) of an aggregate lead to huge variations in the dust production: the resulting dust production is approximately proportional to the effective particle size, 2) the decrease of the filling factor leads not only to an increase of the effective particle size, but also dramatically increases the dust mass production rate due to the changes of the optical characteristics ($Q_{sca}$ and $p$). For a given $Af\rho$ value, the estimated mass production for carbon particles may change from about five ($R_m$=1.6 micron) to twenty ($R_m$=0.16 micron) times.

Results obtained for the scattering particles (silicate in our work) demonstrate a different behaviour. In this case the mass production depends on both the phase function and the scattering efficiency $Q_{sca}$ which is a strongly non-linear function (for details see Appendix), and dust production $M_d$ varies not only with change of particle mass, but also with the change of monomer size. The general behaviour of the curves presented in the right column is more complicated than the one considered above. For example, the difference between the BA and BAM2 aggregates of the same mass is not about the constant (as it was in case of carbon particles). If the monomer size is 0.32 microns, the more compact (BAM2) particles with the mass of M >$4\times10^{-13}$ kg imply higher dust production for the fixed $Af\rho$ value. A similar effect is observed for the aggregates built by the biggest monomer (bottom row). It is interesting to note that the absolute maximum of dust production changes only by factor 4—5 when the aggregates of mass M=$4\times10^{-15}$ kg and M=$5\times10^{-11}$ kg are tested. It means that we cannot reproduce the conclusions formulated above for the carbon particles, and there is no simple monotonic dependence between the dust production and the aggregate size and/or its porosity. We can only conclude that the production may change almost a hundredfold for the porous aggregates in the considered mass interval. Therefore, we can assume that for the transparent particles the accounting of the particle size distribution, as well as the constrains for the monomer size, play an important role. Otherwise, the transformation of the measured $Af\rho$ value into the dust mass production becomes a very speculative and ambiguous procedure that does not have a solid physical basis.

As we noted above, application of the EMT leads to a significant increase of variability. Thus, for the big aggregates the dust production varies tenfold. It means that the size and structure of dust grains can dramatically change the theoretical evaluation of the dust production based on the observed $Af\rho$ parameter. This conclusion is valid for all the tested monomer sizes (different columns).

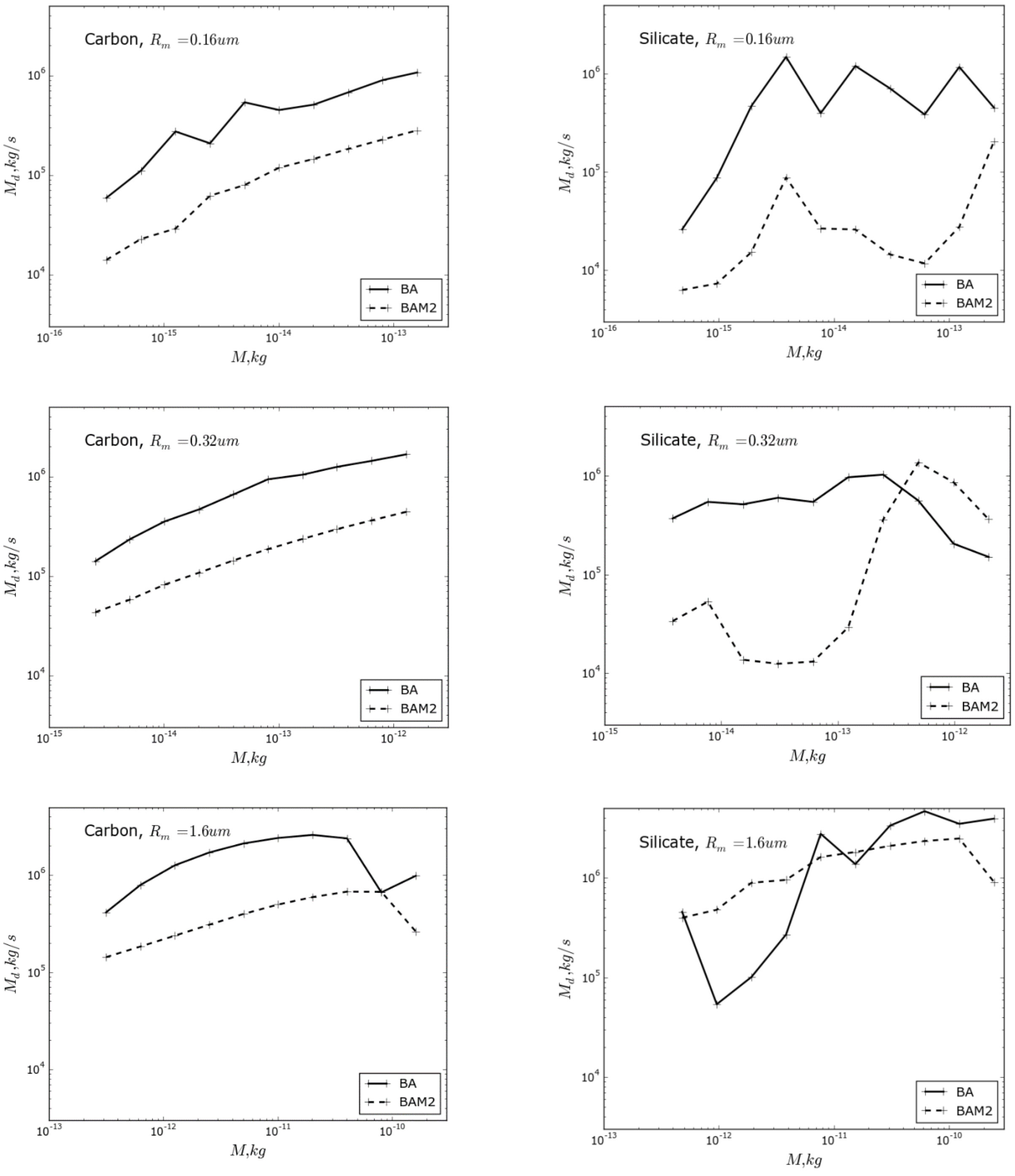

**Fig. 6.** The calculated dust production versus mass of the dust aggregates. The left column: carbon aggregates, the right column: silicate aggregates. Results are shown for the three monomer sizes and two types of aggregates (BA and BAM2).

## Discussion and Conclusions

An Oort cloud comet C/2012 S1 (ISON) passed at only 0.012 AU from the Sun on November 28, 2013 (Knight et al., 2013). The cometary nucleus did not survive the close approach to the Sun, and was completely split (Sekanina and Kracht, 2014). As most of the new comets, which first arrive to the inner part of the Solar System, the ISON comet was active and bright at large heliocentric distances on the pre-perihelion arc. The comet ISON became an interesting object for monitoring of its activity and study physical properties at wide range of heliocentric distances due to the high level of the cometary activity at such a large distance (near 6.3 AU), and predicted extreme proximity to the Sun at perihelion. During its motion in the perihelion passage, the comet demonstrated an outburst activity including an unexpected decrease in activity at around 4.2 AU, and a new strong outburst of activity at about 0.65 AU. We observed the ISON comet at relatively large heliocentric distances, from 6.2 to 4.8 AU. This allowed us to obtain the images of the comet, and analyse how its morphology and dust production change with time. The photometric images derived with the broadband filters from September 29 to December 16, 2012 showed that the comet had a small coma for the first observations, and an extended coma with a small tail on December, 2012 (Fig. 2). The comet tail observation confirms the results reported in (Meech et al., 2013). The brightness of ISON slowly increased during September – October 2012. In December 2012, the brightness of the comet was a few times higher than over the previous months. Our data is well complemented with the observational results presented in (Meech et al., 2013). The authors proposed that the light curve, which covered the heliocentric distances from 9.3 to 4.2 AU is likely controlled by the sublimation of $CO_2$. They also infer that there was a long slow increase in activity beginning at 8.8 AU and peaking at 5 AU, at which point the activity decreased again at 3.5 AU. Since no molecular emissions were detected in the spectra of this comet (at close heliocentric distances of about 4.8 AU, than our photometrical data were obtained), it allows us to study the reflective dust properties without the gas emission effect. The analysis of spectral data shows the growth of dust grain reflectivity with increasing wavelength. The derived reddening equals to 9.3% ± 1.1% per 1000 Å in the range 3800−5400 Å and 2.5% ± 1.2% for the range 5400-6800 Å, respectively. Our results from photometry, obtained at larger heliocentric distances show that the colours of the comet are much redder than the solar ones. The reflectivity gradient of the comet obtained from the photometric data varies from 2.5 to 5.47 % per 1000 Å, which is close to the result obtained for the near nucleus area for this comet at the heliocentric distance of about 4 AU (Li eat al., 2013; Zubko et al., 2015). Our evaluation of the $Af\rho$ value corresponds to the constant dust activity trend of the ISON comet at the distances from 6.2 to 4.8 AU.

Our simulations clearly indicate that to retrieve dust production from the observational $Af\rho$ parameter is an ambiguous task. The result of such a procedure is strongly depended on the dynamical (e.g. effective density and cross-section) as well as optical (e.g. scattering coefficient and phase function) characteristics of dust grains. A variation of the mentioned parameters can lead to dramatic changes in the evaluation of mass production. We demonstrate that the dynamic and the optical properties are interconnected via the microscopic properties of dust grains (effective size and porosity). We also strongly recommend applying the effective medium approach together with the Mie scattering theory to calculate the optical properties of porous dust grains. The other uncertainty in results arises due to the unknown chemical composition of the dust. If there is no additional information about the dust substance, the estimation of mass production becomes highly speculative. Thus, even for the well-known dust sizes the optical properties can vary by an order of magnitude when carbon or silicate particles are tested.

When the variability of dust production is considered, one has to take into account another important factor. Namely, the particle size distribution function. This function is widely used in publications related to the photometry of comets. A simple power function with an exponent varying near −3 is usually used as a model function (Fink and Rubin, 2012). In this paper, we intentionally did not introduce such a model function in order to show the role of basic model parameters more clearly. For the opaque particles (in this paper, these are carbonates), the dependence of dust production on the size of the aggregate has a simple monotonic character, which weakly depends on the size of the monomer. In the case of silicates, all the considered model characteristics (monomer size, aggregate porosity and mass) affect the dust production in a complex nonlinear manner. Obviously, for such particles the choice of the size distribution function will introduce a considerable uncertainty in the estimation of dust production. The last important conclusion relates to the possible presence of large massive dust particles in the coma. Such particles can give a negligible contribution to the luminosity of the coma. However, recent theoretical and observational results have shown that accounting for such large "hidden" particles can reverse the picture. Gundlach et al. (2015) showed that the large grains (up to a few centimeters in size) can be released by CO sublimation at heliocentric distances about 4 AU, based on the theory that comets form by the gravitational instability of an ensemble of dust and ice aggregates (Skorov and Blum, 2012). Taking into account this possibility, together with the finding that a significant (if not nearly all) mass of cometary dust is contained in such large particles (Fulle, 2016), we conclude that the estimation of dust mass production based on the $Af\rho$ parameter is a very dubious undertaking.

## Acknowledgments

The observations at the 6-m BTA telescope were carried out owing to the financial support of the Ministry of Education and Science of the Russian Federation (agreement No. 14.619.21.0004, project ID RFMEFI61914X0004). Collaborative work was done help to the support through the grant DAAD. O. Ivanova thanks the SASPRO Programme, the People Programme (Marie Curie Actions) European Union's Seventh Framework Programme under the REA grant agreement No. 609427, and the Slovak Academy of Sciences for financial support. Research has been further co-funded by the Slovak Academy of Sciences grant VEGA 2/0023/18. Yu. Skorov thanks the Deutsche Forschungsgemeinschaft (DFG) for support under grant SK 264/2-1. A grateful acknowledgment is made to Dr. Uwe Keller for interesting discussion and constructive criticism, and also to Dr. S.M. Andrievsky, Dr. Zakhozhay and Dr. Agapitov for their kind help.

## References

A'Hearn M.F., Schleicher, D.G., Millis R.L., Feldman P.D., Thompson, D.T. 1984. Comet Bowell 1980b. *Astron. J.* 89, 579–591.

Afanasiev, V. L., & Moiseev, A. V. 2011. Scorpio on the 6m telescope: current state and perspectives for spectroscopy of galactic and extragalactic objects. *Open Astron.*, 20, 363-370.

Allen C.W. 1976. *Astrophys. Quant*. *Athlone Press*, *London*. p. 689.

Bodewits D., Farnham T., A'Hearn M. F. 2013. Comet C/2012 S1 (Ison). *CBET*. 3608, 1.

Curdt W., Boehnhardt H., Vincent J. B., Solanki S. K., Schühle U., Teriaca L. 2014. Scattered Lyman-α radiation of comet 2012/S1 (ISON) observed by SUMER/SOHO. *Astron. Astrophys*. 567, L1.

Cochran A., Barker E., Cochran W. 1980. Spectrophotometric observations of P/Schwassmann-Wachmann 1 during outburst. *Astron. J.* 85, 474–477.

DiSanti M., Bonev B., Gibb E., Villanueva G., Paganini L., Mumma M., ...McKay A. 2014. The chemical composition of comet C/2012 S1 (ISON) between 1.2 au and 0.35 au from the Sun. *Asteroids*, *Comets*, *Meteors 2014*.

Draine B. T., Salpeter E. E. 1979. On the physics of dust grains in hot gas. *Astrophys. J.* 231, 77-94.

Green D. 2012. Comet C/2012 S1 (ISON). *CBET* 3238.

Feldman P., McCandliss S., Weaver H., Fleming B., Redwine K., Li M., ...Moseley S. 2014. Far-ultraviolet observations of comet C/2012 S1 (ISON) with a sounding-rocket-borne instrument. *Asteroids, Comets, Meteors 2014*.

Fink U., Rubin M. 2012. The calculation of Afρ and mass loss rate for comets. *Icarus*. 221, 2, 721–734.

Fulle M., Marzari F., Della Corte V., Fornasier S., Sierks H., Rotundi A., et al. 2016. Evolution of the dust size distribution of comet 67P/Churyumov–Gerasimenko from 2.2 AU to perihelion. *The Astrophys. J., 821(1), 19.*

Gundlach B., Blum J., Keller H. U., Skorov Y. V. 2015. What drives the dust activity of comet 67P/Churyumov-Gerasimenko? *Astron. Astrophys., 583, A12.*

Haser, L. 1957. Distribution d'intensité dans la tête d'une comète. Bulletin de la Societe Royale des Sciences de Liege, 43, 740-750.

Iseli M., Küppers M., Benz W., Bochsler P. 2002. Sungrazing comets: Properties of nuclei and in situ detectability of cometary ions at 1 AU. *Icarus*, 155(2), 350-364.

Ivanova O. V., Luk'yanyk I. V., Kiselev N. N., Afanasiev V. L., Picazzio E., Cavichia, O., ... Andrievsky S. M. 2016. Photometric and spectroscopic analysis of Comet 29P/Schwassmann-Wachmann 1 activity. *Planetary Space Science*. 121, 10-17.

Jewitt, D. (2015). Color systematics of comets and related bodies. The Astronomical Journal, 150(6), 201.

Jockers K. 1997. Observations of scattered light from cometary dust and their interpretation. *Earth, Moon, and Planets*, 79(1-3), 221-245.

Jones, G. H., Knight, M. M., Battams, K., Boice, D. C., Brown, J., Giordano, S., ... & Fitzsimmons, A. (2018). The Science of Sungrazers, Sunskirters, and Other Near-Sun Comets. *Space Science Reviews*, *214*(1), 20.

Kartasheva T.A., Chunakova N.M., 1978. Spectral atmospheric transparency at the Special Astrophysical Observatory of the USSR Academy of Science from 1974 to 1976. *Astrof. Issled. Izv. Spets*. 10, 44–51 (Rus.).

Keane J., Milam S., Coulson I., Gicquel A., Meech K., Yang B., Riesen T., Remijan A., Villanueva G., Corrinder M., Charnley S., Mumma M. 2014. Thermal emission from large solid particles in the coma of comet C/2012 S1 (ISON) around perihelion. *Asteroids, Comets, Meteors 2014*.

Kelley M. S., Harker D. E., Wooden D. H., Woodward C. E. 2007. Crystalline Silicates and the Spectacular Comet C/2006 P1 (McNaught). *Bull. of the American Astronomical Society*, 39, 827.

Knight M. M., A'HearnLi M. F., Biesecker D. A., Faury G., Hamilton D. P., Lamy P., Llebaria, A. 2010. Photometric study of the Kreutz comets observed by SOHO from 1996 to 2005. *The Astron. J.*, 139(3), 926.

Knight M. M., Walsh K. J. (2013). Will comet ISON (C/2012 S1) survive perihelion? *The Astrophys. J. Let., 776(1), L5.*

Knight M. M., Battams K. 2014. Preliminary analysis of SOHO/STEREO observations of sungrazing comet ISON (C/2012 S1) around perihelion. *Astrophys. J. Let*. 782(2), L37.

Knight, M. M., & Schleicher, D. G. (2014). Observations of comet ISON (C/2012 S1) from Lowell Observatory. *The Astronomical Journal, 149*(1), 19.

Korsun P.P., Ivanova O.V., Afanasiev V.L. 2006. Cometary activity of distant object C/2002 VQ94 (LINEAR). *Astron. Astrophys*. 459, 977–980.

Korsun P.P., Ivanova O.V., Afanasiev V.L. 2008. C/2002 VQ94 (LINEAR) and 29P/Schwassmann–Wachmann 1 – CO+ and N2+ rich comets. *Icarus*. 198, 465–471.

Korsun P. P., Kulyk I., Ivanova O. V., Zakhozhay O. V., Afanasiev V. L., Sergeev A. V., Velichko, S. F. 2016. Optical spectrophotometric monitoring of comet C/2006 W3 (Christensen) before perihelion. *Astron. Astrophys, 596, A48.*

Kolokolova L., Hanner M. S., Levasseur-Regourd A. C., Gustafson B. A. S. 2004. Physical properties of cometary dust from light scattering and thermal emission. *Comets II*. 577, 184.

Kulyk, I., Jockers, K., Credner, T., & Bonev, T. (2004). The wavelength dependence of the monochromatic extinction coefficient for the observatory on Terskol Peak. *Kinematika i Fizika Nebesnykh Tel*, *20*, 372-378.
Kulyk I., Rousselo, P., Korsun P. P., Afanasiev V. L., Sergeev A. V., Velichko S. F. 2018. Physical activity of the selected nearly isotropic comets with perihelia at large heliocentric distance. *Astron. Astrophys, 611, A32.*
Lamy P.L., Pedersen H., Vio R. 1988. The Dust Tail of Comet P/Halley in April 1986. Exploration of Halley's Comet. *Springer, Berlin, Heidelberg,* p. 661–664.
Lamy, P. L., Toth, I., & Weaver, H. A. 2014. Hubble Space Telescope Observations of the Nucleus of Comet C/2012 S1 (ISON). *TheAstrophys. J.Lett., 794(1), L9.*
Landsman W.B. 1993. The IDL astronomy user's library.
Landolt A. U. 1992. UBVRI photometric standard stars in the magnitude range 11.5-16.0 around the celestial equator. *Astron. J*. 104, 340-371.
Li J. Y., Kelley M. S., Knight M. M., Farnham T. L., Weaver H. A., A'Hearn M. F., ...Toth I. 2013. Characterizing the Dust Coma of Comet C/2012 S1 (ISON) at 4.15 AU from the Sun. *Astrophys. J. Let.* 779(1), L3.
McKay A., Cochran A., Dello Russo A., Weaver H., Vervack R., Harris W., Kawakita H., DiSanti, M., Chanover N., Tsvetanov Z. 2014. Evolution of fragment-species production in comet C/2012 S1 (ISON) from 1.6 au to 0.4 au. *Asteroids, Comets, Meteors 2014*.
Meech K. J., Pittichova J., Bar-Nun, A. Notesco G., Laufer D., Hainaut, O. R., ...Pitts M. 2009. Activity of comets at large heliocentric distances pre-perihelion. *Icarus*,201(2), 719-739.
Meech K. J., Yang B., Kleyna J., Ansdell M., Chiang H. F., Hainaut O., ..Wainscoat R. 2013. Outgassing Behavior of C/2012 S1 (ISON) from 2011 September to 2013 June. *Astrophys. Let*. 776, 2, L20.
Moreno F., Pozuelos F., Aceituno F., Casanova V., Duffard R., Molina A., ...Segundo A. S. 2014. On the Dust Environment of Comet C/2012 S1 (ISON) from 12 AU Pre-perihelion to the End of its Activity around Perihelion. *Astrophys. J.* 791(2), 118.
Nakamura R., Kitada Y., Mukai T. 1994. Gas drag forces on fractal aggregates. *Planetary Space Science.* 42(9), 721-726.
Neckel H., Labs D. 1984. The solar radiation between 3300 and 12,500 Å. *Sol. Phys.* 90, 205–258.
Nevski B., Novichonok A. 2012, *CBET* 3238.
Oke J.B., 1990. Faint spectrophotometric standard stars. *Astron. J*. 99, 1621–1631.
Ootsubo T., Usui F., Takita S., Watanabe J., Yanamandra-Fisher P., Honda M., Kawakita, H. Furusho R. 2014. Mid-infrared observations of sungrazing comet C/2012 S1 (ISON) with the Subaru Telescope. *Asteroids, Comets, Meteors 2014.*
Knight, M. M., & Walsh, K. J. 2013. Will comet ISON (C/2012 S1) survive perihelion?. The *Astrophys. J. Let*., 776(1), L5.
Petrova E. V., Jockers K., Kiselev N. N. 2000. Light scattering by aggregates with sizes comparable to the wavelength: an application to cometary dust. *Icarus*. 148(2), 526-536.
Russo N. D., Vervack R. J., Kawakita H., Cochran A., McKay A. J., Harris W. M., ... Biver, N. 2016. The compositional evolution of C/2012 S1 (ISON) from ground-based high-resolution infrared spectroscopy as part of a worldwide observing campaign. *Icarus*. 266, 152-172.
Rousselot P., Korsun P. P., Kulyk I. V., Afanasiev V. L., Ivanova O. V., Sergeev A. V., Velichko, S. F. 2014. Monitoring of the cometary activity of distant comet C/2006 S3 (LONEOS). *Astron. Astrophys, 571, A73.*
Samarasinha N. H., Mueller B. E., Knight M. M., Farnham T. L., Briol, J., Brosch N., ... & Hergenrother, C. 2015. Results from the worldwide coma morphology campaign for comet ISON (C/2012 S1). Planetary and Space Science, 118, 127-137.
Schleicher, D. G. 2010. The fluorescence efficiencies of the CN violet bands in comets. The *Astron. J., 140(4), 973.*

Sekanina Z. 2003. Erosion model for the sungrazing comets observed with the Solar and Heliospheric Observatory. *Astrophys. J*. 597(2), 1237.

Sekanina Z., Kracht R. 2014. Disintegration of Comet C/2012 S1 (ISON) Shortly Before Perihelion: Evidence from Independent Data Sets. *arXiv preprint arXiv:1404.5968*.

Shen Yue., Draine B. T., Johnson E.T. 2008. Modeling Porous Dust Grains with Ballistic Aggregates. I. Geometry and Optical Properties. *The Astrophys. J.* 689 (1), 260-275.

Scarmato T. 2014. Sungrazer Comet C/2012 S1 (ISON): Curve of light, nucleus size, rotation and peculiar structures in the coma and tail. *arXiv preprint arXiv:1405.3112*.

Skorov Y. V., Keller H. U., Rodin A. V. 2008. Optical properties of aerosols in Titan's atmosphere. *Planetary Space Science, 56(5), 660-668*.

Skorov Y., Blum J. 2012. Dust release and tensile strength of the non-volatile layer of cometary nuclei. *Icarus, 221(1), 1-11*.

Skorov Y., Reshetnyk V., Lacerda P., Hartogh P., Blum, J. 2016. Acceleration of cometary dust near the nucleus: application to 67P/Churyumov-Gerasimenko. *Mon. Not.R. Astron. Soc***.** 461, 3410-3420.

Trigo-Rodríguez J. M., Meech K. J., Rodriguez D., Sanchez A., Lacruz J., Riesen T. E. 2013. Post-Discovery Photometric Follow-Up of Sungrazing Comet C/2012 S1 ISON. *LPI Contributions*, 1719, 1576.

Weaver H., A'Hearn M., Feldman P., Bodewits D., Combi M., Dello Russo N., McCandliss S. 2014. Ultraviolet spectroscopy of comet ISON (2012 S1). *Asteroids, Comets, Meteors 2014*.

Williams G.V. 2013. Comet P/2012 S1 (ISON). *MPEC 2013*, V07.

Wooden D. H., De Buizer J. M., Kelley M. S., Woodward C. E., Harker D. E., Reach W. T., ... Kolokolova L. 2014. Comet C/2012 S1 (ISON)'s carbon-rich and micron-size-dominated coma dust. *Asteroids, Comets, Meteors 2014*.

Woodward C. E., Jones T. J., Brown B., Ryan E. L., Krejny M., Kolokolova L., ...Sitko M. L. 2011. Dust in comet C/2007 N3 (Lulin). *Astron. J*. 141(6), 181.

Zubko E., Muinonen K., Videen G., Kiselev N. N. 2014. Dust in Comet C/1975 V1 (West). *Mon. Not.R. Astron. Soc***.** 440(4), 2928-2943.

Zubko E., Videen G., Hines D. C., Shkuratov Y., Kaydash V., Muinonen K., ...Wooden D. H. 2015. Comet C/2012 S1 (ISON) coma composition at~ 4au from HST observations. *Planetary Space Science*. 118, 138-163.

## Appendix. A model for the optical properties of porous grains.

As we noted in section 4.4, the quantitative evaluation of the $Af\rho$ parameter should be based on the individual optical properties of dust grains. At present, various computer models for calculating these characteristics have been developed (see, for example, Kolokolova, 2004). Nevertheless, the simplest model using the Mie theory approximation continues to be used in the many studies. This popularity is explained by the simplicity of these models. One of the basic simplifications of this approximation is the assumption of the spherical shape of the scattering particles. As shown in (Shen et al, 2008) this idealization is satisfactorily satisfied for a wide class of particles formed as a result of ballistic aggregation. However, within the framework of this approach, it is impossible to take into account the grain porosity. A simple, satisfactory way to do this is to add Maxwell-Garnet effective medium approximation for calculation of the optical characteristics. An account of the grain porosity (both in calculating the effective size and optical characteristics) is an important enhancement to compare with the analysis presented in Fink and Rubin (2012). Hereafter we present such a practical tool that allows us to keep "ease of use" and at the same time to make the next natural step in improving the accuracy of modeling.

Our model grains are formed by ballistic aggregation of indivisible spheres (monomers) which stick at the point of contact within a growing cluster (aggregate). For clarity we assume that the spheres are monodisperse. Structures formed in this way, known as ballistic particle-cluster aggregation (BPCA). Below we test two specific subclasses of BPCA introduced in (Shen et al., 2008): the BA (Ballistic Agglomeration) corresponds to the classical BPCA (see, e.g., Skorov et al., 2008) and results in the most porous random ballistic aggregates; the BAM2 (Ballistic Agglomeration with 2 Migrations) allows for two migrations of the monomer-projectile and produces aggregates of higher density. Detailed information about generation procedures and geometrical properties of the aggregates is presented in the cited work. For our study it is important that the aggregate effective porosity (see Eq. 12 in Shen et al. 2008) is a function of the aggregate mass (or number of monomers). The number of monomers in the aggregate $N_m$ varies from 8 to 4096. Optical characteristics of the aggregates, which consist of a different number of spherical monomers, are examined using the data from the available library of samples (https://www.astro.princeton.edu/~draine/agglom.html). Typical aggregate asphericity (deviation from unit ratio of minimum and maximum projection area) is between 20-30% for BA and 10-15% for BAM2. We consider the resulting aggregate as a sphere of the effective radius equals to the radius of a circle with an area equals to the average aggregate cross-section (Nakamura et al. 1994):

$$R_{eff} = \sqrt{\frac{\sigma}{\pi}},$$

where $\sigma$ is the orientation-averaged cross-section of the aggregate obtained by the averaging of the cross-sections from different projections. The radius of the constituent monomer is taken to be equal to 0.16 microns, 0.32 microns and 1.6 microns. These sizes are selected as the characteristic sizes at which the optical properties of the sphere change significantly in the visible radiation range in the framework of the Mie theory (see Fig. 3 in Fink and Rubin 2012). Silicate and carbon grains with known refraction coefficients (n=1.687+i0.03006 for silicate and n=2.14+i0.805 for carbon) are used. The balk densities are 3.5 g·cm$^{-3}$ and 2.3 g·cm$^{-3}$ for silicate and carbon grains respectively (Draine and Salpeter, 1979). In order to demonstrate the role of grain porosity, a solid sphere with the same radius $R_{eff}$ and constant refraction coefficient is considered as a test particle. The Maxwell-Garnet effective medium approximation is used for the calculation of the optical characteristics, such as scattering efficiency $Q_{sca}$ and phase function $p$.

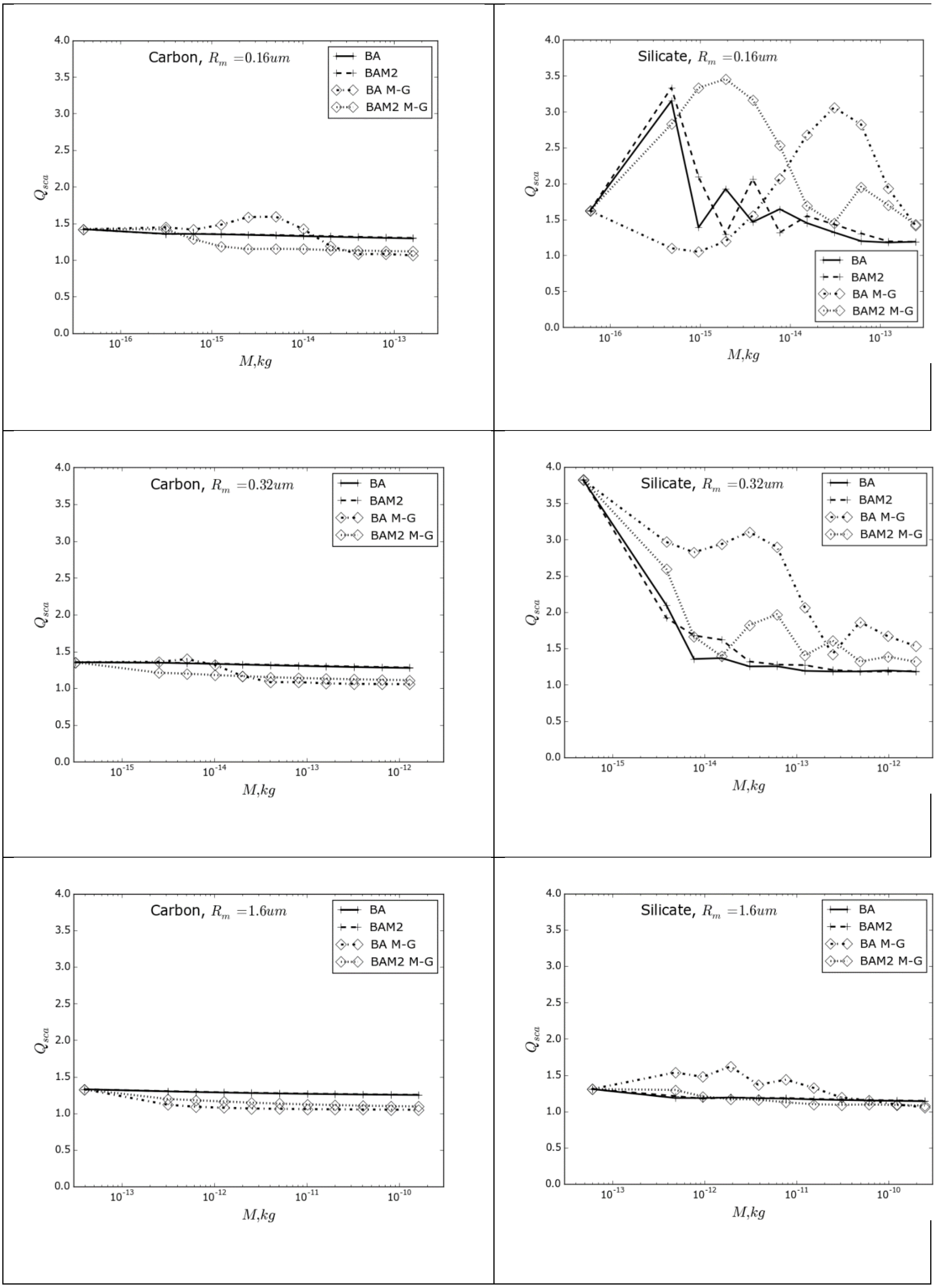

**Fig. A1.** The scattering efficiency $Q_{sca}$ of the dust particle as a function of the aggregate mass *M*. The results are shown for three monomer sizes (different rows), two sorts of grains (carbon – left column, silicates –right column). We examined two types of aggregates (BA and BAM2) and two optical models: with constant refractive index (Mie theory) and refractive index from EMT in the Maxwell-Garnet approximation.

The scattering efficiency of the dust particle $Q_{sca}$ as a function of the aggregate mass $M$ is shown in Fig. A1. We present the results obtained for three monomer sizes, two types of aggregates (BA and BAM2), and two sorts of material (silicate and carbon). In order to prove the importance of using the effective medium theory (EMT), the results obtained for this approximation and for the classical Mie theory are plotted. The wavelength of incident radiation is 0.65 micron which is close to filter R at the phase angle $\alpha$=8.5°. All presented evaluations are executed for the phase angle $\alpha$=8.5° corresponding to our observations of the comet C/2012 S1 (ISON) at heliocentric distances of more than 6 AU.

The left column of Fig. A1 shows the results obtained for the carbon particles consisting of monomers of different sizes. The right column shows the similar results obtained for the silicate particles having the same geometrical properties. In all cases, the number of monomers varies from 1 to 4096, which means that the maximum mass of particles is about $10^{-13}$, $10^{-12}$, and $10^{-10}$ kg, if $R_m$ equals 0.16, 0.32 and 1.6 micron, respectively. We note that the filling factor (percent of the summary monomers volume in the total aggregate volume) of BAM2 particles is approximately two times larger for small particles ($N_m$~10), and three times larger for the particles of the maximum mass ($N_m$ =4096) than the corresponding filling factor of the more porous BA particles. This leads to the fact that even when the simplest Mie scattering model is applied, the difference between the grains of the same mass (but with different effective sizes) is manifested. Since in the Mie theory the refractive index of a homogeneous spherical particle is assumed to be constant, the difference in results (BA vs. BAM2) is due to the difference in size only. In this case, the main variations of the scattering coefficient $Q_{sca}$ arise due to the well-known oscillations from the interference of transmitted and diffracted waves. These oscillations are well detected only for the transparent silicate grains, because absorption reduces the interference oscillations. For the absorbing carbon particles, the oscillations are not observed for all the tested monomer sizes and types of aggregates. The growth of the monomer size (from top to bottom) effectively operates in a similar style damping the oscillations. Hence, the $Q_{sca}$ non-monotony is clearly visible only for the silicate aggregates consisting of the smallest monomers ($R_m$ =0.16 micron). In all cases the $Q_{sca}$ goes to unit with the growth of aggregate effective size. Thus, the difference between the BA and BAM2 aggregates practically disappearsand the effective scattering factor loses its sensitivity to both the chemical composition (carbon vs silicate) and the microstructure (BA vs BAM2) of the aggregates.

When the effective medium theory is applied, the effective refractive index varies with particle structure based on the Maxwell-Garnett mixing rule. In these cases porosity variations lead to some changes in optical properties and, hence, the behaviour of $Q_{sca}$ differs for the BA and BAM2 aggregates of the same mass. This effect is more pronounced than the variations due to the size change in the Mie model: now the characteristics of the BA and BAM2 particles of the same mass differ by hundreds ($R_m$ =0.16 micron) or tens ($R_m$ =1.6 micron) of percent. Note that this is true in the case when the monomer size is the smallest (top row), the scattering factor varies most strongly when the aggregates of different microstructure are examined. We conclude that the use of the effective medium theory approximation is important for the calculation of the optical properties based on the reasonable arguments in favour for the assumption fact that the cometary porous particles contain monomers with size 0.1-0.2 micron (Petrova et al., 2000, Kolokolova et al., 2004; Skorov et al., 2016).

The values of the scattering phase function $p$ evaluated for the considered scattering angle as a function of the aggregate mass $M$ are plotted in Fig. A2. The figure structure is the same as the structure of Fig. A1: we examine carbon and silicate aggregates of two classes (BA and BAM2) consisting of differently sized monomers ($R_m$ equals 0.16, 0.32 and 1.6 microns). As in the case of the scattering coefficient, the phase curves calculated for the carbon particles have a smoother behaviour. Small oscillations are observed only for the smallest monomers (the upper left plot). In the entire range of monomer sizes, the phase functions calculated in the effective medium approximation differ markedly from the functions calculated in the Mie scattering model. In the latter case, the phase functions change little with the increasing mass, and only for the largest particles constructed from the largest monomers (the lower left plot) there is a relative growth of scattering function with the aggregate mass. Similar increases of the phase function are observed for these particles when the EMT is applied, although the absolute values of $p$ are about an order of magnitude smaller for this model. The more porous BA aggregates of the same mass are more transparent, and their phase function $p$ above the functions calculated for the BAM2 aggregates. In the Mie scattering model the results are practically independent on the aggregate porosity. Thus, we conclude that for the absorbing grains (left column) the phase scattering function strongly depends on the model approximations, and the EMT model should be used for our porous aggregates. In the case of much more transparent material (right column) the behaviour of the scattering functions becomes more complicated. For the fine aggregates consisting of smallest monomers, conspicuous oscillations are observed in all the tested models (right top panel). At the same time, the character of oscillations is different for the Mie and the EMT scattering models: while in the Mie scattering model the differences between BA and BAM2 for large particles decrease, in the EMT scattering model the differences are most noticeable in this mass range (M>$10^{-14}$ kg). When the bigger monomer size is used (middle and bottom panels) the difference between the BA and BAM2 phase scattering functions decreases in the Mie scattering model, but remains well pronounced in the EMT scattering model. We recall that our calculations are performed for one wavelength and for one value of the scattering angle. Therefore, our results are presented in the form which is slightly different from the generally accepted one: we show how the optical characteristics of particles ($Q_{sca}$ and $p$) change as their mass grows and how their porosity and the size of the base monomer change.

The results presented above clearly show that the optical characteristics of dust aggregates should be evaluated in the combination of the Mie scattering theory with the Maxwell-Garnett mixing rule explicitly taking into account the aggregate porosity. We can expect that this development of the model will play a significant role in the estimation of the dust production based on the observed classical $Af\rho$ parameter.

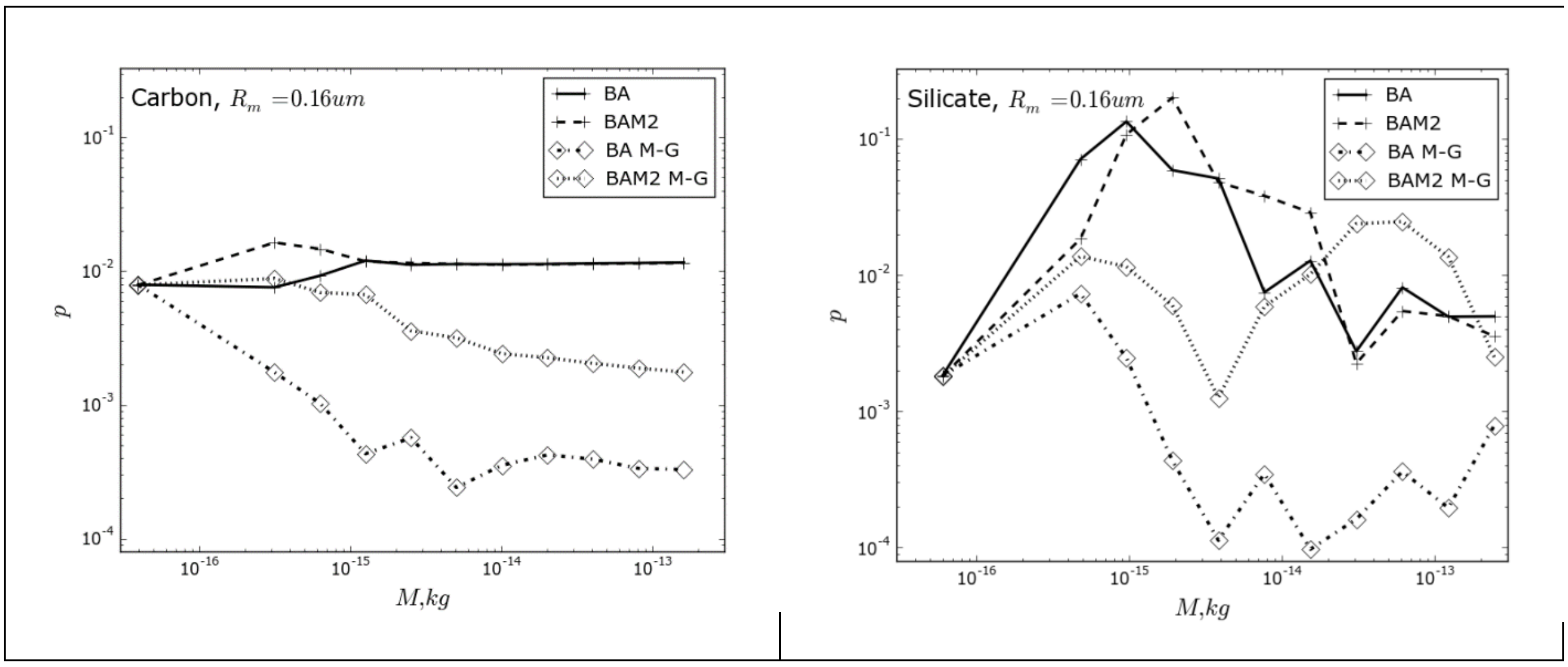

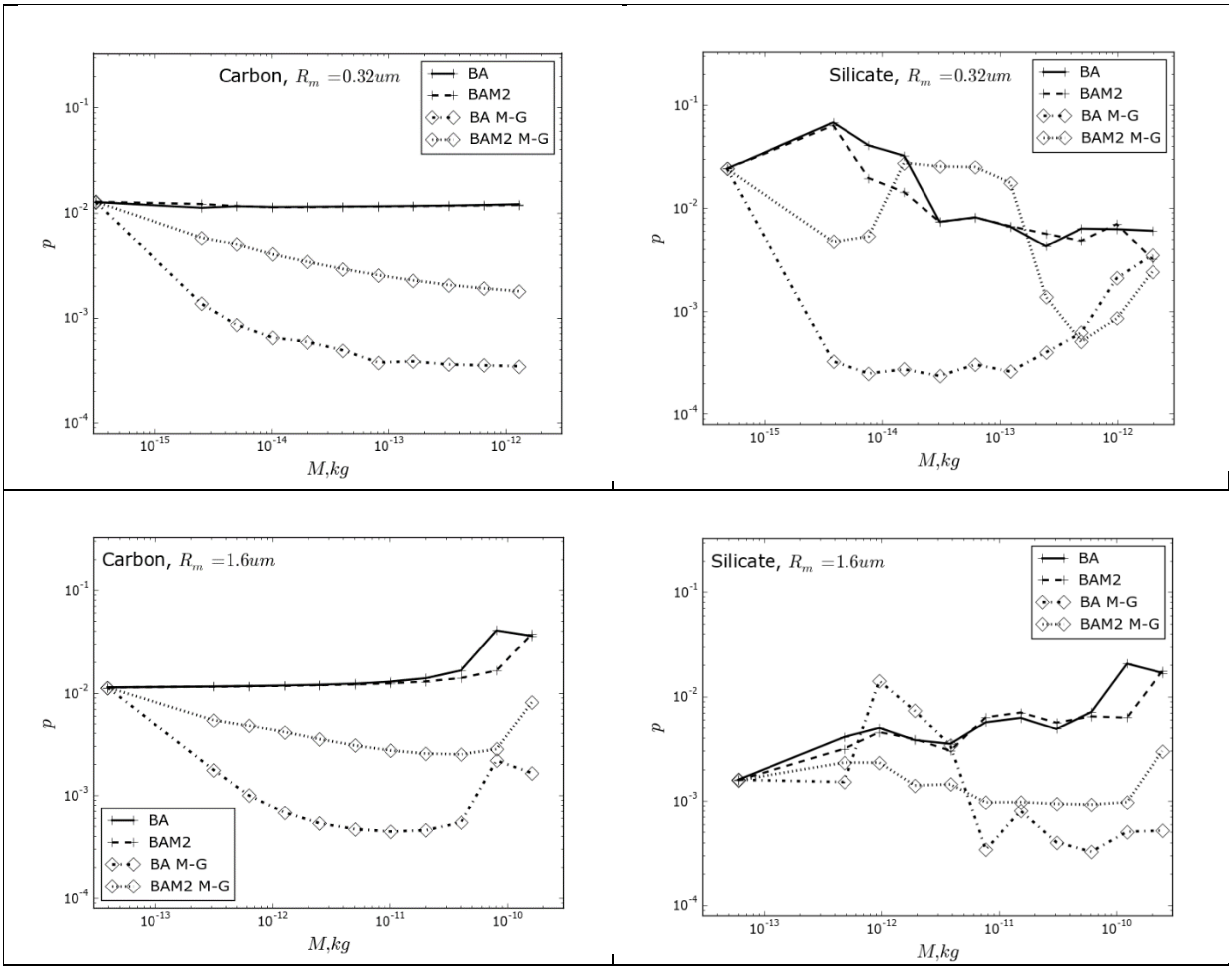

**Fig. A2.** The scattering phase function *p* evaluated for one angle (8.5 degree) as a function of the aggregate mass *M*. The results are shown for three monomer sizes (different rows), two sorts of grains (carbon – left column, silicates –right column). We examined two types of aggregates (BA and BAM2) and two optical models: with constant refractive index (Mie theory) and refractive index from EMT in the Maxwell-Garnet approximation.